\documentclass[aps,
preprint,
superscriptaddress]{revtex4}

\usepackage{amsfonts}
\usepackage{amsmath}
\usepackage{amssymb}
\usepackage{graphicx}%
\usepackage{hyperref}
\usepackage[dvipsnames]{xcolor}
\usepackage{ulem}

\begin{document}

\title{Coulomb correlations and magnetic properties of L1$_0$ 
FeCo: \\ a DFT+DMFT study}



\author{A. S. Belozerov}
\affiliation{M. N. Miheev Institute of Metal Physics, Russian Academy of Sciences, 620108 Yekaterinburg, Russia}

\author{A. A. Katanin}
\affiliation{Center for Photonics and 2D Materials, Moscow Institute of Physics and Technology, 141701 Dolgoprudny, Russia}
\affiliation{M. N. Miheev Institute of Metal Physics, Russian Academy of Sciences, 620108 Yekaterinburg, Russia}
\affiliation{Skolkovo Institute of Science and Technology, 121205 Moscow, Russia}

\author{V. I. Anisimov}
\affiliation{M. N. Miheev Institute of Metal Physics, Russian Academy of Sciences, 620108 Yekaterinburg, Russia}
\affiliation{Skolkovo Institute of Science and Technology, 121205 Moscow, Russia}
\affiliation{Ural Federal University, 620002 Yekaterinburg, Russia}


\begin{abstract}
We consider electronic correlation effects and their impact on magnetic properties of tetragonally distorted chemically ordered FeCo alloys (L1$_0$ structure) being a promising candidate for rare-earth-free permanent magnets. We employ a state-of-the-art method combining density functional and dynamical mean-field theory. According to our results, the predicted Curie temperature reduces with increase of lattice parameters ratio $c/a$ and reaches nearly 850~K at ${c/a=1.22}$. For all considered $c/a$ from 1 to $\sqrt{2}$, we find well-localized magnetic moments on Fe sites, which are formed due to strong correlations originating from Hund's coupling. At the same time, magnetism of Co sites is more itinerant with a much less lifetime of local magnetic moments. However, these short-lived local moments are also formed due to Hund's exchange. Electronic states at Fe sites are characterized by a non-quasiparticle form of self-energies, while the ones for Co sites are found to have a Fermi-liquid-like shape with quasiparticle mass enhancement factor ${m^*/m\sim 1.4}$, corresponding to moderately correlated metal. The strong electron correlations on Fe sites leading to Hund's metal behaviour can be explained by peculiarities of the density of states, which has pronounced peaks near the Fermi level, while weaker many-body effects on Co sites can be caused by stronger deviation from half-filling of their $3d$ states. The obtained momentum dependence of magnetic susceptibility suggests that the ferromagnetic ordering is the most favourable one except for the near vicinity of the fcc structure and the magnetic exchange is expected to be of RKKY type.
\end{abstract}

\maketitle

\section{Introduction}
There are several key magnetic properties a permanent magnet should possess for high-performance industrial applications.
These properties are connected with high values of saturation magnetization, Curie temperature, coercivity,
and
magnetocrystalline anisotropy energy (MAE).
In widely used Sm-Co and Nd-Fe-B magnets, high magnetization and Curie temperature are mainly provided by Fe or Co constituents, while large magnetic anisotropy is due to rare-earth elements with strong spin-orbit coupling. 

For rare-earth-free permanent magnets, there is a need in another source of large magnetocrystalline anisotropy.
Such a source can be provided, e.g., by tetragonal distortion in $L1_0$ structure of AuCu-type with atomic monolayers alternating along the $c$ axis.
The well-known examples are {$L1_0$} FePt, MnAl, and FeNi, where magnetic anisotropy constants are close to those of rare-earth-based magnets~\cite{Klemmer1995,Lewis2014}.

Another example is $L1_0$ FeCo, which was predicted to possess highly desirable characteristics for permanent magnets (for review, see Ref.~\cite{Cui2018}),
%
namely, a large uniaxial magnetocrystalline anisotropy of 10~MJ/m$^3$ and saturation magnetization of 2.2~$\mu_{\rm B}$/atom were obtained by Burkert \textit{et al.}~\cite{Burkert} within density functional theory (DFT) at lattice constants ratio ${c/a = 1.22}$. 
This ratio corresponds to the body-centered tetragonal (bct) structure, which is almost equally distant from the bcc (${c/a = 1}$) and fcc (${c/a = \sqrt{2}}$) lattices.
In addition to chemically ordered $L1_0$ structures, large MAE was also predicted for disordered Fe$_{1-x}$Co$_x$ alloys at $x$ about 0.5--0.65 and $c/a$ about 1.2--1.25~\cite{Burkert}.

Although bulk samples of $L1_0$ FeCo have not been fabricated yet, the tetragonal distortion in FeCo was obtained in epitaxially grown layers on Pd~\cite{Winkelmann2006,Yildiz2009_1,Yildiz2009_2},
Ir~\cite{Yildiz2009_1,Yildiz2009_2}, Rh~\cite{Luo2007,Yildiz2009_1,Yildiz2009_2,Oomiya2015},
Pt~\cite{Moulas2008},
and Cu$_3$Au~\cite{Ohtsuki2014,Ponce2018} substrates.
Tetragonal Fe-Co alloys were also grown as a constituent of Fe$_{0.36}$Co$_{0.64}$/Pt
superlattices~\cite{Andersson2006,Warnicke2007}, where a huge perpendicular MAE, reaching 210~$\mu$eV/atom, and a saturation magnetization of 2.5~$\mu_{\rm B}$/atom at 40~K were measured~\cite{Andersson2006}.
Nanopatterned FeCo layers were fabricated by
Hasegawa \textit{et al.}, who reported a perpendicular uniaxial magnetic anisotropy of 2.1~MJ/m$^3$ and a coercivity of 0.6~T~\cite{Hasegawa2017}.
In addition to films, Gong \textit{et al.} grew a FeCo shell on fcc AuCu core, which both were then transformed into tetragonal structure~\cite{Gong2017}.

Another approach to stabilize the tetragonal distortion in Fe-Co is to use interstitial doping
with light elements such as C~\cite{Delczeg-Czirjak2014}, N~\cite{Odkhuu2019}, or B~\cite{Reichel2014,Reichel2015}.
%
There was also an attempt by Gao \textit{et al.} to incorporate tungsten with large spin-orbit coupling in Fe-Co films, which resulted in large magnetization
and enhanced perpendicular coercive fields of 2–3~kOe at low W concentration~\cite{Gao2013}.

Previous theoretical studies of tetragonal Fe-Co systems were performed within DFT.
These  studies  addressed  the  origin of large MAE in thin films~\cite{Schonecker2016}, superlattices~\cite{Andersson2006,Hyodo2015} and bulk samples~\cite{Burkert,Kota2014,Steiner2016,Turek2012,Turek2013},
as well as its dependence on chemical composition~\cite{Burkert,Andersson2006,Schonecker2016,Delczeg-Czirjak2014,Reichel2014,Reichel2015,Reichel2017,Odkhuu2019} and chemical order~\cite{Neise2011,Schonecker2016,Turek2012}.
A mechanism of large MAE was proposed by Burkert \textit{et al.}, who
showed that it can be caused by peculiarities of electronic states near the Fermi level~\cite{Burkert}.
Moreover, the chemically ordered Fe–Co films were found by Neise \textit{et al.} to have a much larger magnetic anisotropy than the disordered ones~\cite{Neise2011}.
A strong reduction of the Curie temperature with increase of $c/a$ was obtained by Jakobsson \textit{et al.} by mapping the DFT results onto the classical Heisenberg model~\cite{Jakobsson2013}.
In addition, the effect of interstitial doping on magnetic anisotropy and structural stability of tetragonal Fe-Co was studied~\cite{Delczeg-Czirjak2014,Reichel2014,Reichel2015,Reichel2017,Odkhuu2019}.

%

Partially filled $3d$ subshells in Fe and Co ions may result in significant many-body effects. The treatment of these effects in DFT calculations can be improved, e.g., by avoiding symmetry restrictions~\cite{DFT_symmetry_restrictions}, employing a sophisticated exchange-correlation functional, and/or combining with a disordered local moment (DLM) method~\cite{dlm_method} to simulate a paramagnetic state. An accurate treatment of correlation effects can also be achieved by combining DFT with model approaches, which are usually based on Heisenberg-like or Hubbard models.

In material specific calculations the Hubbard model is often solved using the static mean-field approximation (DFT+$U$ method~\cite{Anisimov1991}) or dynamical mean-field theory (DMFT)~\cite{dmft}. The latter explicitly takes into account the temporal quantum correlations and thermal fluctuations, and becomes exact in the limit of infinite coordination number. Capturing of local spin dynamics within DMFT approach (supplemented by its combination with DFT method ~\cite{dftdmft} for description of realistic materials) becomes
especially relevant for studying the (partial) {\it formation} of local magnetic moments \cite{spin_freezing,Hund_metals,Stadler,Deng,MazitovKatanin} and origin of finite temperature metallic magnetism. By means of the DFT+DMFT approach, important information about magnetic and structural properties of iron~\cite{alpha_iron2010,OurAlphaIgoshevKatanin,OurAlphaBelozerovKatanin,leonov_fe_Lichtenstein2001_Katsnelson_Grechnev_Benea2012,Hausoel,Minar2005} and its alloys~\cite{Hausoel,Minar2005,Leedahl} was obtained.
An essential role of Coulomb correlations in the B2 structure of FeCo, that has no tetragonal distortion, was recently shown using DFT+DMFT method~\cite{Gerasimov}.

In the present paper, we employ the DFT+DMFT method to study the interplay of electronic and magnetic properties in $L1_0$ FeCo, as well as to analyze persistence of local magnetic moments.
Since the presence of long-range magnetic order hides the local magnetic properties, we enforce the paramagnetic state by assuming spin-independent self-energy
in all our calculations, except those of uniform magnetic susceptibility. This allows us to get insight into intrinsic properties of the magnetically ordered phase by investigation of electronic and dominating magnetic correlations.
We consider various tetragonal distortions in the $L1_0$ structure, including the limiting cases of bcc and fcc lattices.

\section{Method and computational details} \label{sec:computational_details}

We perform our study using a fully charge self-consistent DFT+DMFT approach~\cite{charge_sc} implemented with plane-wave pseudopotentials~\cite{espresso,Leonov1}.
Vanderbilt ultrasoft pseudopotentials
with the Perdew-Burke-Ernzerhof form of the generalized gradient approximation were used.
The lattice constants were taken from a previous DFT study~\cite{Burkert}, where the largest MAE was found at ${c/a=1.22}$ with ${a = 2.683}$~\AA\, and ${c = 3.273}$~\AA.
{
The thermal expansion of the lattice was neglected in our calculations. However, we checked that considering thermal expansion by using equilibrium unit cell volume leads to qualitatively similar results.
}
The convergence threshold for total energy was set to $10^{-5}$~Ry.
The kinetic energy cutoff for wavefunctions was set to 70~Ry.
The integration in the reciprocal space was carried out using ${16\times 16\times 16}$\ $\textbf{k}$-point mesh in all calculations except those of the momentum-dependent susceptibility, where ${50\times 50\times 50}$ mesh was employed.
Our DFT+DMFT calculations explicitly include the $3d$, $4s$ and $4p$ valence states of Fe and Co, by constructing a basis set of atomic-centered Wannier functions {(not maximally localized)} within the energy window spanned by the $s$-$p$-$d$ band complex~\cite{Wannier}.

We parametrize the Coulomb interaction in the $3d$ shell via Slater integrals $F^0$, $F^2$, and $F^4$ linked to the Hubbard parameter ${U\equiv F^0}$ and Hund's rule coupling ${J_{\rm H}\equiv (F^2+F^4)/14}$. 
%
In our calculations we adopt ${U=4}$~eV and ${J_{\rm H}=0.9}$~eV for both Fe and Co.
These values are in agreement with estimates for elemental iron~\cite{Belozerov2014} and were widely used in its DFT+DMFT studies~\cite{OurAlphaBelozerovKatanin,OurGammaKataninBelozerov,OurAlphaGammaTransition}.
We also checked that considering ${U=3}$~eV
does not qualitatively affect our results. 
%
%
To account for the electronic interactions described by DFT, we use the around mean-field form of double-counting correction, evaluated from the self-consistently determined local occupations.
We also verified that the fully localized form of double-counting correction leads to similar results. 
The impurity problem in DMFT was solved by the hybridization expansion continuous-time quantum Monte Carlo method~\cite{CT-QMC} with the density-density form of Coulomb interaction.
To compute the density of states, we perform the analytical continuation of self-energies from imaginary to real frequencies by using Pad\'e approximants~\cite{Pade}.
%
\section{Results and discussion}
\subsection{Electronic properties\label{Sect:elProp}}

\begin{figure}[t]
\centering
\includegraphics[clip=true,width=0.45\textwidth]{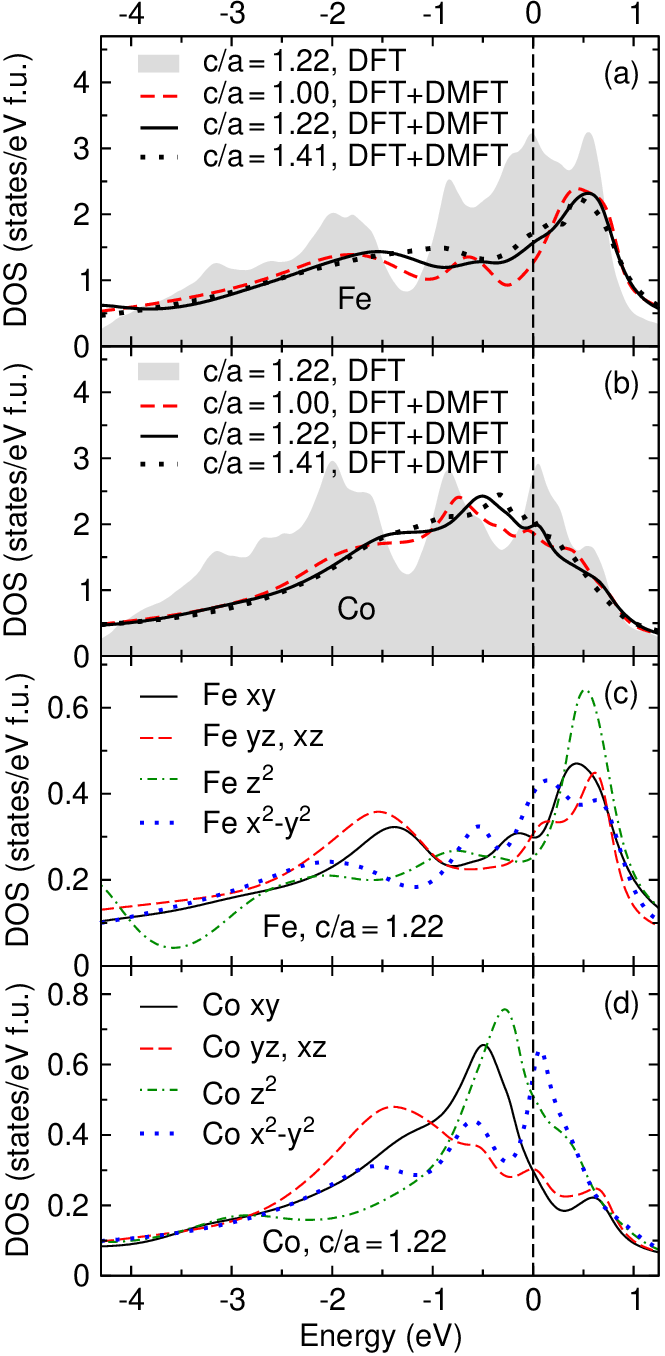}
\caption{
\label{dos}
Total (a, b) and orbital-projected (c, d) density of $3d$ states obtained by non-magnetic DFT (a, b) calculations in comparison with DFT+DMFT (a -- d)
for Fe (a, c) and Co (b, d) sites
at temperature ${T=1160}$~K. 
Panels (c) and (d) correspond to ${c/a = 1.22}$. 
The Fermi level is at zero energy.}
\end{figure}

Our DFT+DMFT calculations at ${c/a=1.22}$ yield the $d$-states filling of 6.35 and 7.47 for Fe and Co sites, respectively.
These values are weakly affected by tetragonal distortion and change less than 0.02 for 
$c/a$ in the range from 1 to $\sqrt{2}$.
We also find the filling of various Fe~$d$ orbitals in the range 1.18--1.33 with that of $z^2$ symmetry being the most close to half-filling, that may enhance correlation effects~\cite{spin_freezing}. At the same time, the partial fillings of Co~$d$ states are much farther from half-filling and vary from 1.40 to 1.57.
{We also note that the full charge self-consistency in the DFT+DMFT scheme is crucially important for description of FeCo as its neglect leads to non-physical redistribution of charge density between Fe and Co sites due to different strength of electronic correlations.}

In Fig.~\ref{dos} we present the density of $3d$ states (DOS)
obtained in non-spin-polarized calculations
at temperature ${T=1160}$~K
($\beta\equiv 1/T = 10$~eV$^{-1}$).
As seen in panels (a) -- (b) of Fig.~\ref{dos},
treating the electronic correlations in DMFT results in a renormalization of DOS 
and its significant suppression near the Fermi level.
One can see that the DOS for both constituents has peaks near the Fermi level,
which are substantially smeared due to temperature effects and electronic correlations.
%
As shown in previous studies of iron and model systems, 
such peaks may significantly enhance the many-body effects 
and lead to the Hund's metal behavior~\cite{alpha_iron2010,BelozerovKataninAsymery2018}.
The most significant peaks in FeCo are from the states of
$z^2$ and ${x^2{-}y^2}$ character, resembling the case of bcc iron,
where they originate from the $e_g$ states~\cite{alpha_iron2010}.
Moreover, Co $z^2$ and ${x^2{-}y^2}$ states have the largest DOS at the Fermi level, that also favors the correlation effects by increasing the number of virtual electron-hole excitations~\cite{BelozerovKataninAsymery2018}.

\begin{figure}[t]
\centering
\includegraphics[clip=true,width=0.54\textwidth]{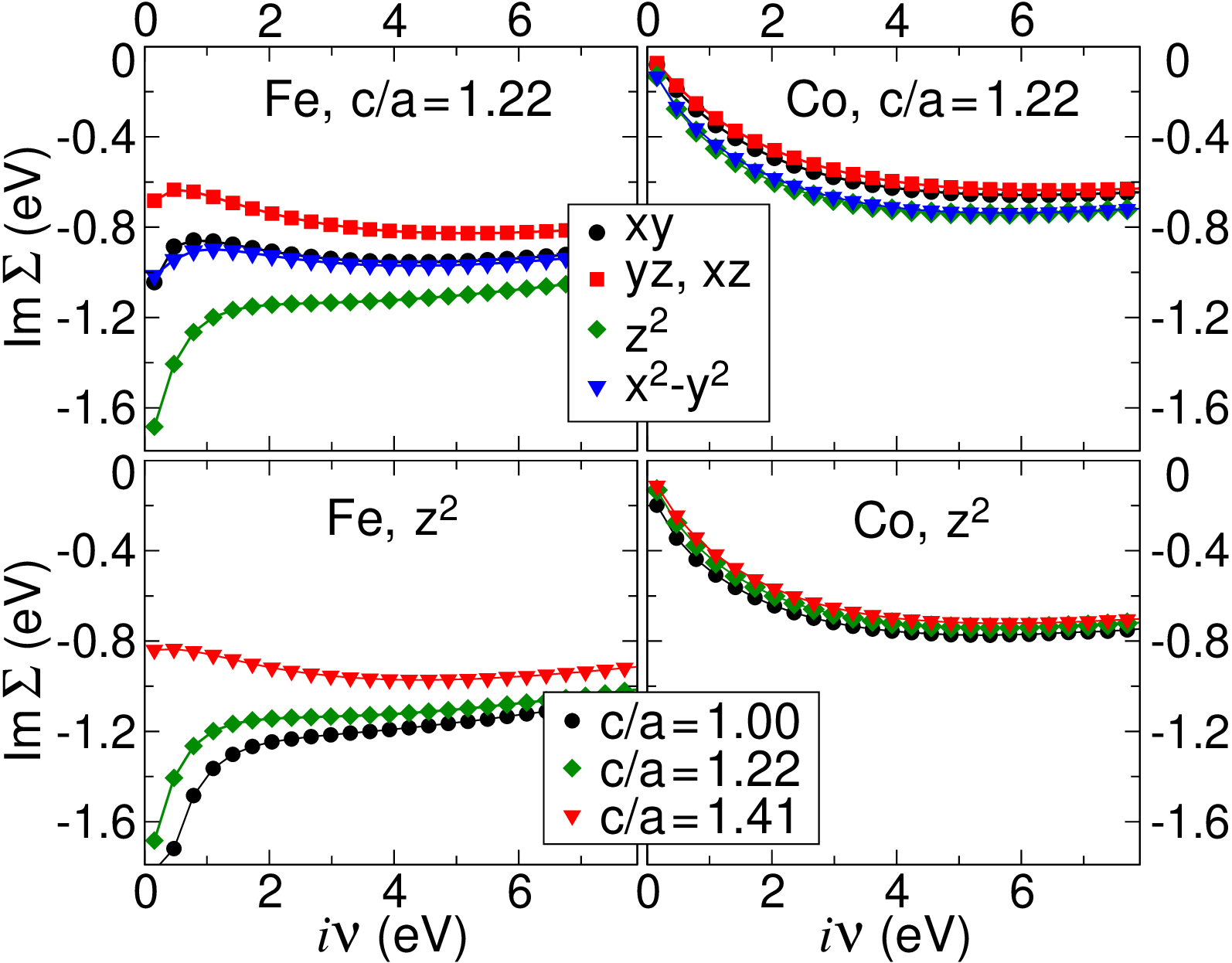}
\caption{
\label{Fig:self_energy}
Imaginary part of electronic self-energy on Fe (left panels) and Co (right panels) sites as a function of imaginary frequency $i\nu$ obtained by DFT+DMFT method at temperature ${T = 580}$~K. The upper panels correspond to ${c/a=1.22}$, while in lower ones the self-energies for $z^2$ states are shown at various $c/a$.}
\end{figure}

To reveal the origin of
DOS suppression at the Fermi level,
in Fig.~\ref{Fig:self_energy} we display the imaginary part of electronic self-energy $\Sigma(i\nu_n)$ as a function of fermionic Matsubara frequency $\nu_n$.
In the presence of well defined quasiparticles, the imaginary part of self-energy depends on small $|\nu_n|$ as ${\textrm{Im}\, \Sigma(i\nu_n)\approx -\Gamma -(Z^{-1} {-}1)\nu_n}$, where $\Gamma$ is the quasiparticle damping (inverse quasiparticle lifetime), which is finite at a finite temperature, and $Z$ is the quasiparticle residue, which is equal in DMFT to the inverse quasiparticle mass enhancement factor ${m/{m^*}}$ due to locality of the self-energy. Since $Z<1$ for well defined quasiparticles, this implies negative derivative $d\,{\rm Im} \Sigma (i\nu)/ d\nu < 0$ at ${\nu\to 0}$, which is accompanied by the minimum of $|{\rm Im}\Sigma(\nu)|$ at the $\nu=0$ along the real frequency axis, corresponding to the minimal scattering rate of quasiparticles at the Fermi surface.

At ${c/a = 1.22}$ the self-energies for all Fe $d$ states
show the non-quasiparticle behavior (the above mentioned derivative of the frequency dependence of the self-energy is positive), implying that interacting electrons in these states cannot be described as Landau quasiparticles with renormalized mass.
This behaviour is similar to that of $e_g$ states in bcc Fe~\cite{alpha_iron2010,OurAlphaIgoshevKatanin,OurAlphaBelozerovKatanin} and is also found at other considered values of $c/a$, but becomes less pronounced when $c/a$ approaches $\sqrt{2}$. In particular, for the limiting case of the fcc lattice (${c/a=\sqrt{2}}$), the non-quasiparticle shape of self-energy is found only for states of $z^2$ character (see Fig.~\ref{Fig:self_energy}). These states, corresponding to the largest peak in Fe DOS near the Fermi level, also show the most non-quasiparticle
shape at other considered values of $c/a$.
To determine the role of Hund's exchange, 
we perform DFT+DMFT calculations with turned off Hund's coupling by setting ${J_{\rm H}=0}$.
In these calculations, all self-energies at ${c/a}$ from 1 to $\sqrt{2}$ are found to have the quasiparticle 
shape (not shown in figure), that indicates an important role played by the Hund's exchange.

At the same time, the behavior of self-energies for Co sites is completely different, 
namely, they have a Fermi-liquid-like form
with small quasiparticle damping, implying a presence of long-lived quasiparticles 
at all considered tetragonal distortions ($c/a\leq \sqrt{2}$). 
{
A similar behavior of self-energies was reported for the B2 structure of FeCo alloy with bcc lattice~\cite{Gerasimov}, that corresponds to ${c/a=1}$ in our notations.
}
To estimate the strength of electronic correlations on Co sites, we calculate 
the quasiparticle mass enhancement factor for each orbital $k$ as
$(m^*/m)_k\,{=}\, 1 - \left[d\,{\rm Im} \Sigma_k (i\nu)/ d\nu \right]_{\nu\to 0}$ and then average over $d$ states of Co. 
The obtained average ${m^*/m}$ is found to increase monotonically from 1.40 to 1.49 as ${c/a}$ grows from 1 to $\sqrt{2}$.
These values characterize Co sites as being moderately correlated.
Our calculations with ${J_{\rm H}=0}$ leads to a drop of $m^*/m$ to 1.25 at ${c/a=1.22}$, indicating that a significant part of electronic correlations on Co sites is also due to Hund's exchange.

\subsection{Magnetic properties\label{Sect:elProp}}

\begin{figure}[b]
\centering
\includegraphics[clip=true,width=0.55\textwidth]{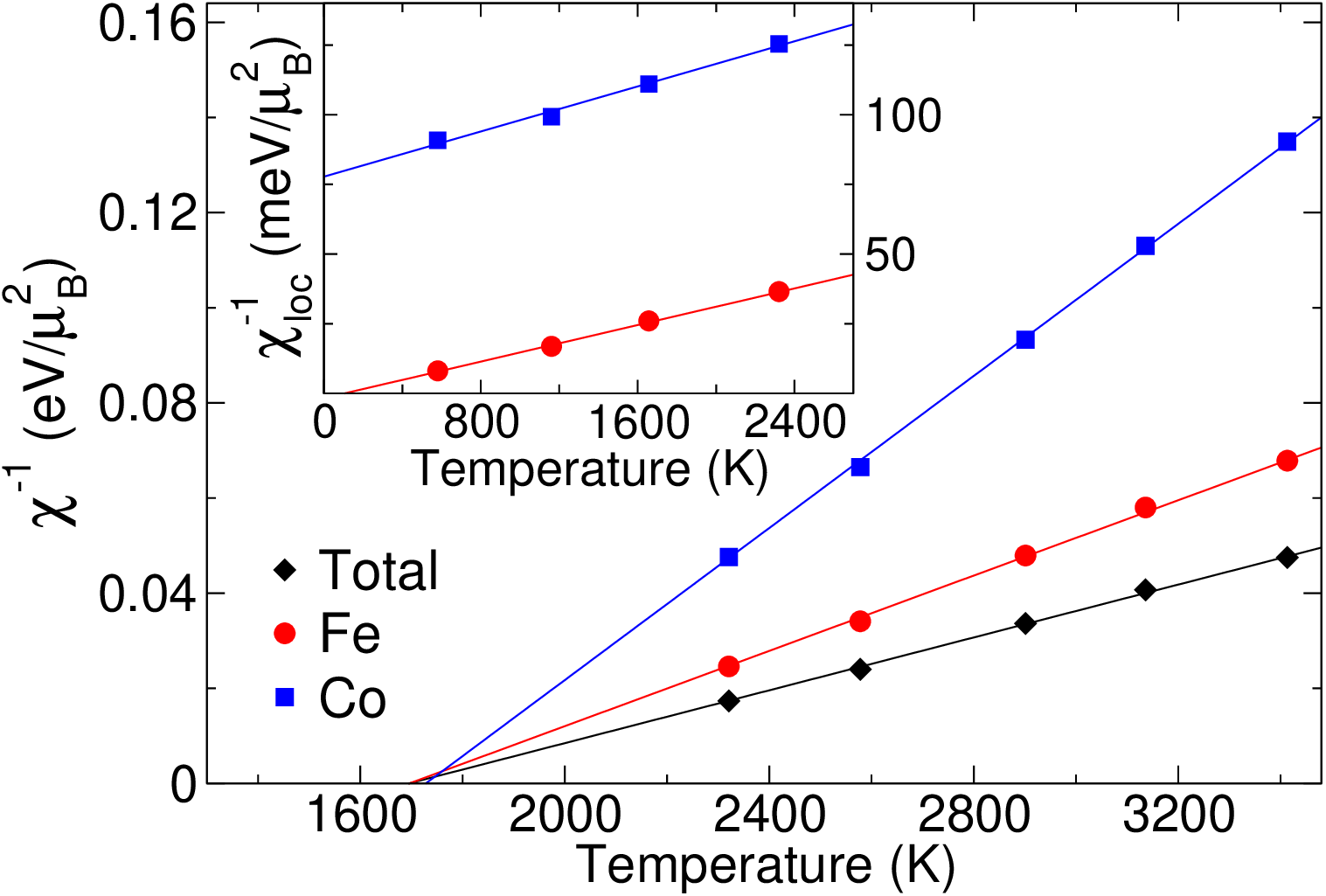}
\caption{
\label{Fig:chi_uniform}
Temperature dependence of inverse uniform (main panel) and local (inset) magnetic susceptibility calculated by DFT+DMFT at ${c/a = 1.22}$. The straight lines depict the least-squares fit to the linear dependence.}
\end{figure}

First, we calculate the uniform magnetic susceptibility as a response to a small external magnetic field, which was checked to provide a linear response. In particular, we use the magnetic field corresponding to splitting of the single-electron energies by 10~meV.
In the main panel of Fig.~\ref{Fig:chi_uniform}, we present the inverse of uniform magnetic susceptibility for the case of ${c/a=1.22}$, where the largest MAE was predicted~\cite{Burkert}. A linear dependence on temperature is clearly seen, which corresponds to the Curie-Weiss law. Our results indicate that the dominant contribution to uniform susceptibility is provided by Fe sites, while the Co contribution is about twice smaller. Extrapolating linearly the inverse susceptibility, we extract the Curie temperature of about 1700~K. In view of the two times overestimation of the Curie temperature in DMFT due to the Ising symmetry of Hund's exchange and mean-field approximation (see Refs.~\cite{Hausoel,BelozerovSU2}), the expected Curie temperature is about 850~K
near ${c/a = 1.22}$~\cite{footnote1},
which is appropriate for technological applications.

\begin{figure}[b]
\centering
\includegraphics[clip=true,width=0.53\textwidth]{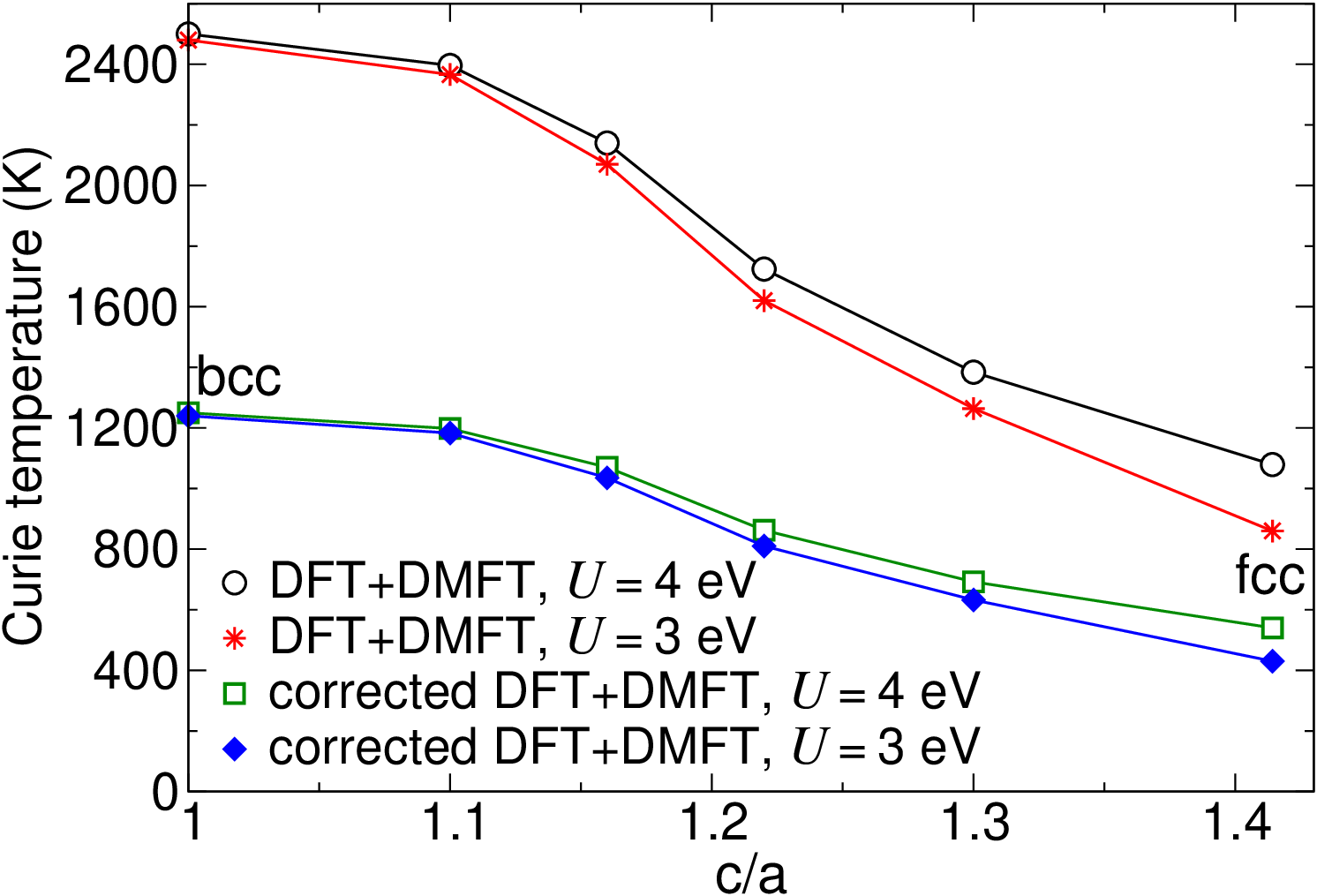}
\caption{\label{fig:curie}
Curie temperature as a function of lattice parameters ratio $c/a$ obtained by DFT+DMFT method with two values of Hubbard $U$.}
\end{figure}

In Fig.~\ref{fig:curie} we show the calculated Curie temperature as a function of lattice parameters ratio
$c/a$. In addition, we also present corrected $T_{\rm C}$ values obtained by division of the former by two due to the described above approximations in DMFT.
One can observe that the $T_{\rm C}$ decreases monotonically with increase of $c/a$,
reaching a maximum of about 1200 K in bcc (${c/a = 1}$) structure.
We have verified that the Hubbard parameter~$U$ weakly affects the value of $T_{\rm C}$, especially far from ${c/a=\sqrt{2}}$ (see Fig.~\ref{fig:curie}). A similar weak dependence of $T_{\rm C}$ on $U$ was reported for bcc Fe~\cite{Belozerov2014}, also exhibiting Hund's metal behaviour~\cite{alpha_iron2010}.
We note that our estimates of $T_{\rm C}$ are about 300~K smaller than those obtained by DFT calculations mapped onto the classical Heisenberg model~\cite{Jakobsson2013}, though the reduction of $T_{\rm C}$ with increase of $c/a$ agrees well in both studies.
{
In addition, our $T_{\rm C}$ values for ${c/a = 1}$ differ less than 50~K from estimates obtained by previous DFT+DMFT study of this structure~\cite{Gerasimov}.
}

{
Next we calculate local magnetic moments in the ferromagnetic state at temperature $T=580$~K. For ${c/a = 1}$ we obtain magnetic moments of 2.92 and 1.83~$\mu_\textrm{B}$ for Fe and Co sites, respectively, which decrease gradually with increase of $c/a$ in good agreement with DFT studies~\cite{Burkert,Jakobsson2013}. In particular, at ${c/a = 1.22}$ our magnetic moments of 2.83 and 1.70~$\mu_\textrm{B}$ agree well with values of about 2.8 and 1.7~$\mu_\textrm{B}$ for Fe and Co sites, respectively, obtained by Jakobsson \textit{et al.} within DFT~\cite{Jakobsson2013}.
}

To investigate the formation of local magnetic moments, we turn off the spin polarization below the calculated Curie temperature and compute local static susceptibility as 
$\chi_{\rm loc}=4\mu_{\rm B}^2 \int_0^\beta \langle S_z(\tau) S_z(0) \rangle d\tau$,
where $S_z$ is the $z$-component of the
local spin operator and $\tau$ is the imaginary time.
The inverse of $\chi_{\rm loc}$ at ${c/a = 1.22}$ is shown in inset of Fig.~\ref{Fig:chi_uniform} and also demonstrates the Curie-Weiss behavior.
In this case, the absolute value of Weiss temperature $T_{\rm loc}$ is proportional to the Kondo temperature $T_{\rm K}$ with the numerical factor of order of unity \cite{Wilson,Melnikov,Tsvelik,Comment,Reply}.
As seen in inset of Fig.~\ref{Fig:chi_uniform},
Fe atoms are characterized by small $T_{\rm K}$, indicating well-formed local magnetic moments.
In contrast, $T_{\rm K}$ for Co atoms is rather large, implying that local moments are not fully formed.
This is in accordance with the Fermi-liquid-like self-energies for Co~$3d$ states. For other $c/a$ from 1 to $\sqrt{2}$, we obtain similar results. 

\begin{figure}[t]
\centering
\includegraphics[clip=true,width=0.53\textwidth]{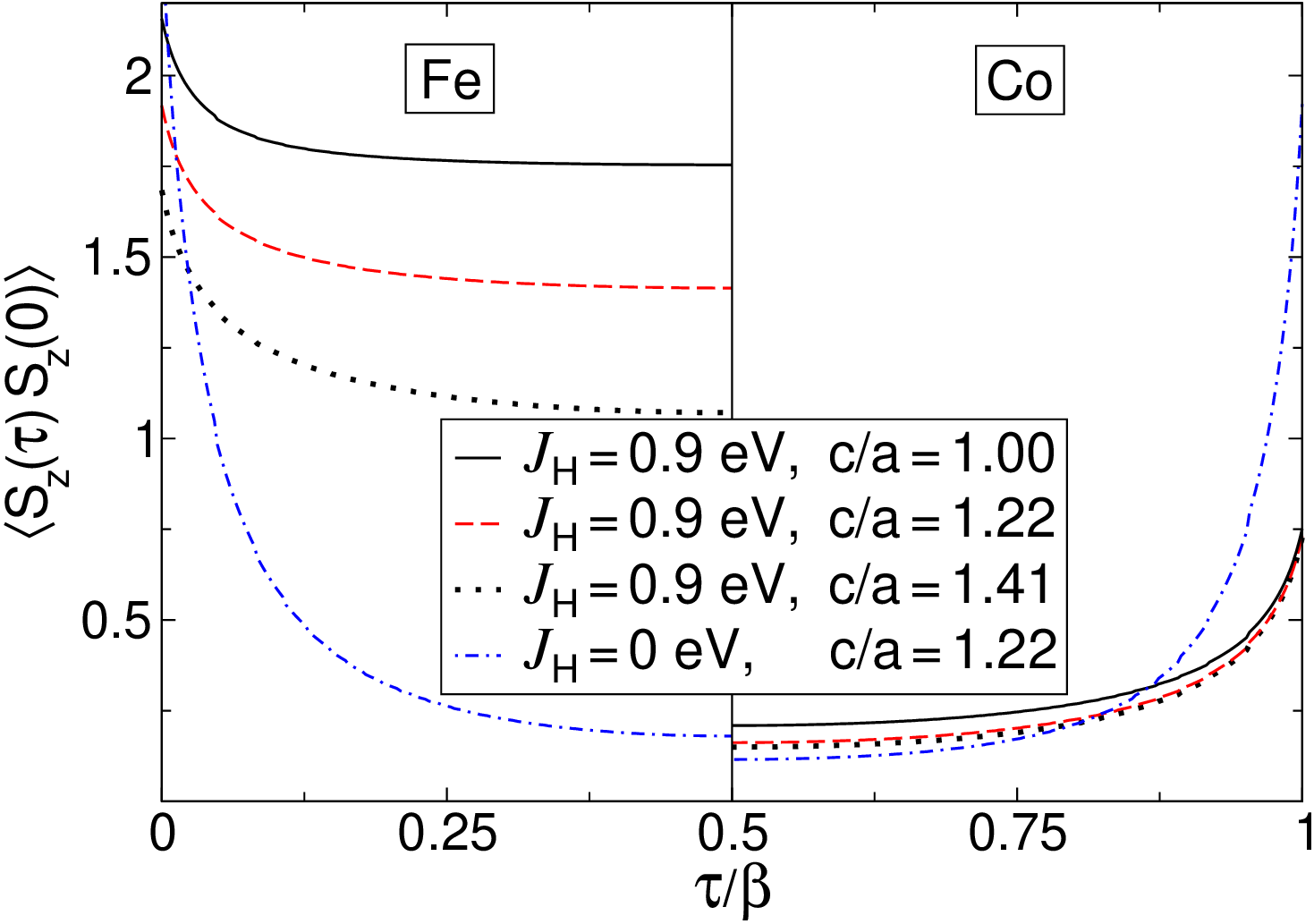}
\caption{
\label{Fig:chi_loc}
Local spin-spin correlation function for Fe (left panel) and Co (right panel) sites in the imaginary-time domain calculated by DFT+DMFT method at 
temperature ${T = 1160}$~K
with various lattice constants ratio $c/a$ and 
Hund's coupling $J_{\rm H}$.}
\end{figure}

\begin{figure}[t]
\centering
\includegraphics[clip=true,width=0.55\textwidth]{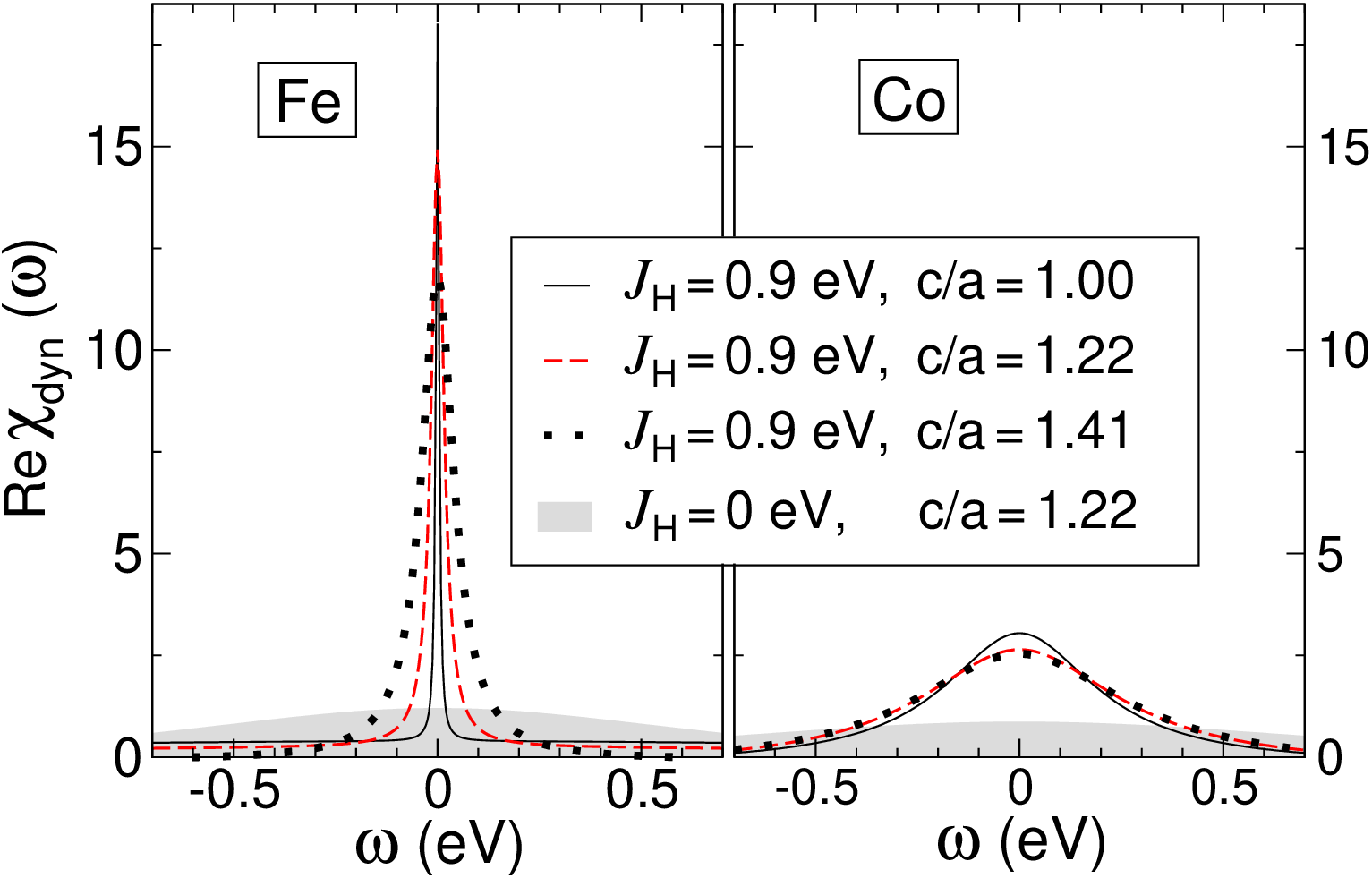}
\caption{
\label{Fig:chi_w}
Real part of dynamical susceptibility for Fe (left panel) and Co (right panel) sites as a function of real-frequency $\omega$ calculated by DFT+DMFT method at 
temperature ${T = 1160}$~K
with various lattice constants ratio $c/a$ and 
Hund's coupling $J_{\rm H}$.}
\end{figure}

In Fig.~\ref{Fig:chi_loc} we show the dependence of local spin-spin correlation function ${\chi_{\rm dyn}(\tau) = \langle S_z(\tau) S_z(0) \rangle}$ on imaginary time $\tau$
with Hund's coupling $J_{\rm H}=0.9$~eV and $J_{\rm H}=0$.
{Since in the latter case ${\chi_{\rm dyn}(\tau)}$ depends weakly on $c/a$, we present results only at ${c/a = 1.22}$.}
One can see that ${\chi_{\rm dyn}(\tau)}$  at ${J_{\rm H}=0}$ has a significant instantaneous average ${\langle S_z^2 \rangle\simeq2}$, corresponding to instantaneous spin $S$ close to~2, and decays rapidly with increase of $\tau$, implying weak localization of magnetic moments for both Fe and Co sites. However, in calculations with $J_{\rm H}=0.9$~eV, ${\chi_{\rm dyn}(\tau)}$ at Fe sites decays slower with $\tau$ than in the case with $J_{\rm H}=0$, and the instantaneous average is almost twice larger for Fe sites than for Co ones. This indicates that Hund's exchange leads to an increase of magnetic moments localization at Fe sites
for all considered $c/a$.

To get a more quantitative estimate of spin localization,
we compute the real-frequency dependence of dynamic susceptibility $\chi_{\rm dyn}(\omega)$, obtained 
by Fourier transform of ${\chi_{\rm dyn}(\tau)}$ to imaginary bosonic frequency and subsequent analytical continuation to real frequency $\omega$ using Pad\'e approximants~\cite{Pade}.
The obtained 
real part of $\chi_{\rm dyn}(\omega)$ is displayed in Fig.~\ref{Fig:chi_w}.
%
The half width of the peak in Re$\chi_{\rm dyn}(\omega)$
at half of its height yields approximately inverse lifetime of local magnetic moments \cite{OurGamma2013,AlessandroTime}.
Therefore, we obtain that Hunds's coupling is responsible for formation of local magnetic moments on Fe sites, while magnetic moments on Co sites are much less localized in line with results on local magnetic susceptibility. Nevertheless, Hund's exchange also contributes substantially to partial formation of local moments at Co sites, similarly to previous results for $\gamma$-iron \cite{OurGamma2013}.
Thus, the peculiarities of magnetic properties of Fe sites can be characterised as the Hund's metal behavior~\cite{spin_freezing,Hund_metals}.

{
We note that relative contributions of Hubbard~$U$ and Hund's coupling~$J_\textrm{H}$ to the origin of correlation effects were addressed in studies of a degenerate three-band Hubbard model~\cite{Stadler}
and two archetypal correlated metals, V$_2$O$_3$ and Sr$_2$RuO$_4$~\cite{Deng,Comment,Reply}.
Results of these studies suggest that, in contrast to Hund's metals, the Mott physics (Hubbard~$U$) becomes dominant in correlated metals, which are in proximity to a Mott insulating state.
}

\begin{figure}[t]
\centering
\includegraphics[clip=true,width=0.55\textwidth]{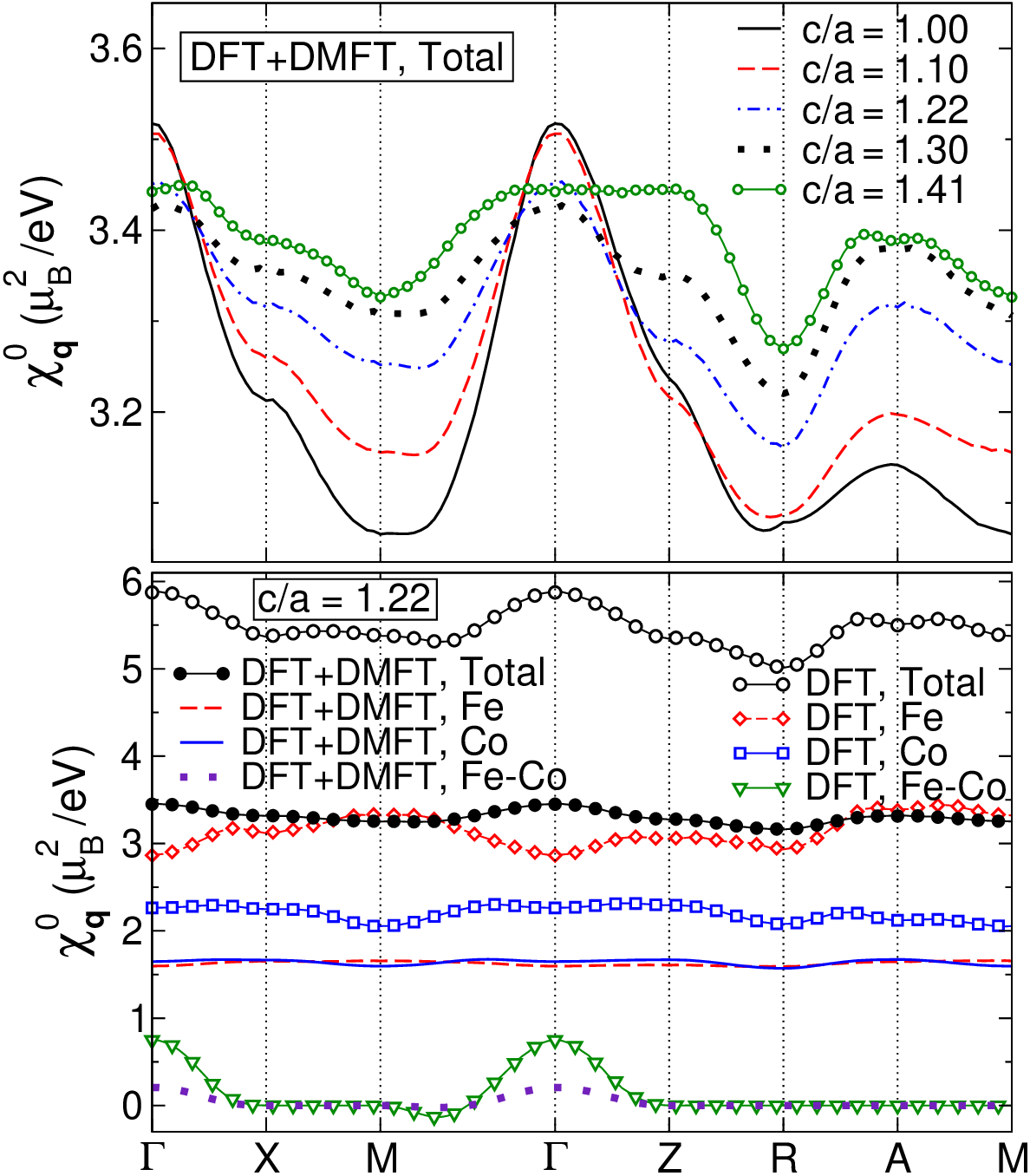}
\caption{\label{fig:chi_q}
Momentum-dependence of the particle-hole bubble
at various lattice parameters ratio $c/a$ (top panel) obtained within DFT+DMFT 
and its partial contributions from Fe and Co sites (bottom panel) obtained within DFT and DFT+DMFT at ${c/a = 1.22}$ and temperature ${T = 1160}$~K.}
\end{figure}

To determine the relative strength of magnetic correlations with various wave vectors,
we compute the irreducible static non-uniform magnetic susceptibility $\chi_{\bf q}^{0}$ as a particle-hole bubble diagram,
\begin{eqnarray}
\chi_{\bf q}^{0} &=& -\frac{2\mu_{\rm B}^2}{\beta} \sum_{{\bf k},\nu_n,i,j,m,m'} G^{im,jm'}_{\bf k} (i\nu_n) G^{jm',im}_{\bf k+q}(i\nu_n),
\end{eqnarray}
where $G^{im,jm'}_{\bf k} (i\nu_n)$ is the one-particle Green's function for $d$ states obtained using the Wannier-projected Hamiltonian at momentum ${\bf k}$,
$\nu_n$ are the fermionic Matsubara frequencies,
$\mu_{\rm B}$ is the Bohr magneton,
$\{i,j\}$ and $\{m,m'\}$ are the site and orbital indexes, respectively.

As seen in the top panel of Fig.~\ref{fig:chi_q}, $\chi_{\bf q}^{0}$ obtained in DFT+DMFT has its global maximum at the $\Gamma$ point (${{\bf q}=}0$) for all considered values of ${c/a < 1.41}$. Therefore, the ferromagnetic ordering is the most favourable one except for the near vicinity of the
fcc structure (${c/a = \sqrt{2}}$). The height of the peak near the $\Gamma$ point decreases with increasing $c/a$, showing that ferromagnetic correlations become less pronounced. 
At the same time, the value of $\chi_{\bf q}^{0}$ at local maximum near point A is significantly less than that at point~$\Gamma$, implying the absence of competing magnetic instabilities.


In the bottom panel of Fig.~\ref{fig:chi_q} we present a partial contribution to $\chi_{\bf q}^{0}$ at ${c/a = 1.22}$. 
One can see that ferromagnetism is favoured in both DFT and {DFT+}DMFT approaches. However, the maximum of $\chi_{\bf q}^{0}$ appears mainly because of the mixed Fe-Co contribution, while contributions from Fe and Co sites are almost momentum-independent in DMFT, similarly to bcc iron~\cite{OurAlphaIgoshevKatanin} and L1$_0$ structure of FeNi~\cite{OurFeNiMagnet}.
We also find the same momentum-dependence of $\chi_{\bf q}^{0}$ at other tetragonal distortions, though the mixed Fe-Co contribution decreases monotonically by a factor of 3 as $c/a$ increases from 1 to $\sqrt{2}$.
Thus,
similarly to L1$_0$ FeNi~\cite{OurFeNiMagnet}, we expect an RKKY type of magnetic exchange between {long-lived} Fe magnetic moments due to virtual hopping between Fe and Co sites.

\section{Conclusion} \label{sec:conclusions}

In summary, we have studied magnetic properties of L1$_0$ FeCo
taking into account Coulomb correlation effects by the DFT+DMFT method.
We find that this prospective candidate to rare-earth-free magnets also possess a quite high Curie temperature at tetragonal distortions far from the fcc structure. In particular, we obtain a Curie temperature estimate of 850~K at lattice parameters ratio ${c/a = 1.22}$, predicted earlier to provide the best MAE~\cite{Burkert}.

Our results indicate that magnetic moments on Fe sites are well localized due to Hund's exchange, which is accompanied by
non-Fermi-liquid behavior of electron self-energy for Fe~$3d$ states.
%
%
At the same time, the magnetism of Co sites is more itinerant with a much less lifetime of local magnetic moments. These short-lived local moments are also formed due to Hund's exchange. However, in contrast to the Fe sites, the self-energies for Co sites have a Fermi-liquid-like shape, resulting in quasiparticle mass enhancement factor $m^*/m\sim 1.4$. This value characterizes Co sites as being moderately correlated.

We find that DOS of Fe and Co atoms has peaks near the Fermi level. Such peaks may significantly enhance the Coulomb correlation effects, and thus affect the magnetic properties, as found previously in bcc iron~\cite{alpha_iron2010} and in model studies~\cite{BelozerovKataninAsymery2018}. At the same time, the Fe sites are found to be much more correlated than Co ones, that may be caused by the proximity of Fe $d$ states to half-filling.

Considering different tetragonal distortions,
we find that the Curie temperature and lifetime of local magnetic moments decrease gradually with increase of $c/a$ from 1 to $\sqrt{2}$. In addition, a competing antiferromagnetic instability appears in the near vicinity of the fcc structure (${c/a=\sqrt{2}}$).
Hence, the tetragonally distorted Fe-Co alloys are expected to show more prominent magnetic characteristics when stabilized at lower values of $c/a$.

The magnetic properties of the L1$_0$ phase of FeCo has much in common with the same phase of FeNi, another promising rare-earth-free magnet~\cite{OurFeNiMagnet}. Although the tetragonal distortion in them is significantly different (L1$_0$ FeNi has a slightly distorted fcc structure with ${c/a=1.424}$), they both show a Hund's metal behavior of Fe sites with well-formed local magnetic moments, while other sites provide more itinerant contribution.
In addition, the analysis of local spin correlation function and momentum-dependent magnetic susceptibility suggests the RKKY-type of magnetic exchange.
Thus, we suppose that in both magnets Fe sites serve as a main source of large well-localized magnetic moments, while the other constituents are required to obtain high magnetic anisotropy associated with coupling of spin and lattice degrees of freedom.
Although we did not account for the effect of lattice vibrations, we expect that it does not change substantially the obtained results, since the correlations, originating from Hund's exchange, already produce substantial broadening of the spectral functions. This broadening is expected to be stronger than the effect of lattice vibrations, similarly to the previous DLM study of Ref. \cite{vibrations}.

The case of L1$_0$~FeCo shows us the feasibility to achieve 
desirable magnetic properties for high-performance permanent magnets using only abundant $3d$ metals.
Therefore, further {theoretical and} experimental efforts, aimed at study and synthesis of systems with tetragonally distorted structure, are of great importance.

\begin{acknowledgments}
The DMFT calculations were supported by the Russian Science Foundation (Project No. 19-72-30043).
The DFT calculations were supported by the Ministry of Science and Higher Education of the Russian Federation (theme “Electron” No. AAAA-A18-118020190098-5).
\end{acknowledgments}



\begin{thebibliography}{99}
\bibitem{footnote1}
Our DFT+DMFT calculations of uniform magnetic susceptibility at equilibrium unit cell volumes yield a Curie temperature of 720~K at $c/a = 1.22$. These equilibrium volumes computed at electronic temperatures from 2300 to 3400 K differ about 3--5\% from that presented in Ref.~\cite{Burkert}.
\end{thebibliography}
\end{document}